# QUANTUM PARTICLES PASSING THROUGH A MATTER-WAVE APERTURE

JIAN-PING PENG

*Department of Physics, Shanghai Jiao Tong University, Shanghai 200240, China*

We investigate theoretically a dilute stream of free quantum particles passing through a macroscopic circular aperture of matter-waves and then moving in a space at a finite temperature, taking into account the dissipative coupling with the environment. The portion of particles captured by the detection screen is studied by varying the distance between the aperture and the screen. Depending on the wavelength, the temperature, and the dimension of the aperture, an unusual local valley-peak structure is found in increasing the distance, in contrast to traditional thinking that it decreases monotonically. The underlying mechanism is the nonlocality in the process of decoherence for an individual particle.

## 1. Introduction

The wave property of a massive particle proposed by Louis de Broglie is an essential ingredient in the development of quantum mechanics. Besides the classical experiments of the diffraction of electrons and neutrons,[1] numerous experiments have been performed recently to demonstrate the property with other elementary particles [2-5] and even with composite nanoparticles, such as fullerenes,[6] He clusters[7] or large fluorinated molecules.[8] Demonstrations of the buildup of quantum interference patterns from individual particles of electrons[9] and fullerenes[10] have shown that both the particle and the quantum wave character can be visualized in one and the same image. Starting with a nonequilibrium many-body wave function, Snoke *et al* have shown the possibilty to deduce the irreversible behavior for a closed system entirely on the basis of wave mechanics without invoking the concept of a random ensemble of particles.[11] Recently, the wave aspect of a single quantum particle has been used to study its finite-time thermodynamics.[12] The key idea is to describe a massive quantum particle in the non-relativistic regime by the wave with a finite wave front instead of the ordinary plane wave with an infinite wave front. Statistics could be performed directly on matter waves associated with the forward-going wave front and those diffracted at its edge. This makes it possible a new mechanism for the thermal interaction with the surround space at a finite temperature. The time-dependent internal energy and the entropy of the particle have been studied for the irreversible process started from a fully coherent quantum state to thermodynamic equilibrium with the surrounding space.[13]

Due to the fact that it is difficult to trace the trajectory of a single quantum particle without interference with its state, here we investigate theoretically a stream of quantum particles with monochromatic wave length passing through a macroscopic circular aperture of matter-waves and are detected in the downstream screen with single particle resolution. The space is set to be at a constant temperature and dissipative coupling with the environment is considered. The particle density in the stream is assumed to be dilute enough so that interaction and correlation between particles are negligible. The theory of single particle thermodynamics is applied to the system, since individual particles in the stream can be believed to move independently within the coherence time. The portion of particles reaching the screen is studied as a function of the distance between the aperture and the screen. Depending on both the wavelength and the temperature, a valley-peak structure is found in increasing the distance of the detection screen, in contrast to traditional thinking that it decreases monotonically. This unusual feature is related to the nonlocality of matter-wave associated with each particle, which is important in the process of decoherence. Experimental verification would be a demonstration of wave-particle duality in a system different from the conventional atom interferometers. Furthermore, information could be obtained to understand the process of dissipative decoherence for an individual quantum particle started from a pure quantum mechanical state.

## 2. Calculations and Results

The system considered is shown schematically in Fig. 1. A stream of quantum particles of mass $m$ and with monochromatic wave length $\lambda$ is described by plane waves traveling in the $x$-direction. Here an individual particle is assumed to be of mass $m$ with no internal structure. A wall at the origin with a hole where particles can pass through acts as the matter-wave aperture. The radius of the hole $a_0$ is set to be much larger than the wave length $\lambda$, so that the diffraction angle is approximately zero and the linear dimension of the forward-going wave-front remains unchanged. In a real experiment with a particle stream, it is required that $a_0$ is small enough to ensure that only a portion of particles from the source can pass through the hole. The particle density in the stream must be dilute enough so that there is very little chance for two wave packets to overlap. A particle entering the aperture carries no temperature associated with its source and effects arising from interaction and correlation are neglected. A downstream screen $D$ apart can detect particles reached with single particle resolution. For electrons, single particle detection is possible using a two-dimensional position-sensitive counting system which is a combination of a fluorescent film and the photo-counting image acquisition system.[9] For molecules of fullerene, particles are captured by a thermally reconstructed Silicon surface, and then molecular deposit is imaged with single particle resolution using scanning tunneling microscopy.[10] The whole space between the aperture and screen is potential free and is set to be at constant temperature $T$. The space here may be filled with electromagnetic radiation just as the cosmic background in the universe.

Will a particle passed through the aperture move forward always at constant speed? In quantum mechanics, it is described by a wave-packet sharply peaked at the de Broglie wavelength $\lambda$ and the wave-packet propagates at group velocity $V_g = h/m\lambda$, with $h$ being the Planck's constant. The time-dependent evolving is governed by the Fresnel-Huygens principle. In contrast to conventional thought of a free quantum particle in which the linear dimension of a forward-going wave-front is implicitly assumed to be infinite large, the radius of the aperture in the present system is finite but much larger than the wavelength. The time-dependent evolution of waves passed through the aperture, strictly speaking, can not be described in a usual Schroedinger formulation of a free particle, and should be described in the Feynman's path integral formulation of quantum theory. Furthermore, a point at the edge of the wave-front is assumed to generate out-going fully spherical waves, besides that a point in the central part of the wave-front generates forward-going semi-spherical waves. As the forward-going plane-wave front travels, diffraction at the edge leads to losing in kinetic energy described by it. As the result, waves diffracted at the edge at different time represent different energy states and their relative probability can be determined. According to the basic idea of quantum physics, the property of an individual particle is determined by matter-waves associated with it. In principle, an individual particle may be in any of states determined by these waves, *i.e.*, the particle itself constitutes automatically a thermodynamic system. If thermal interaction between the particle's system and the surrounding space is allowed, a finite-time thermodynamics of a single particle becomes possible. Without the need to calculate the propagator for a system started from a finite wave-front in a space at a finite temperature, the time dependent probability distribution can be obtained within the framework of equilibrium statistical physics. The correct answer is that the particle will not move forward at constant speed. It absorbs or gives out heat and will evolve continuously from a pure coherent quantum mechanical state into an incoherent thermodynamic state in which it moves equally to all directions. The time scale for the process is of the order $ma_0L/2h$, where an unknown parameter $L$ appears and is assumed to be temperature dependent.[13] In general, the decoherence of a free particle is determined by not only its wavelength but also temperature of the environment and spacial extension of the matter wave associated with it.

Before the wave of a particle hits the detection screen, the partition function for an individual particle is written in the form $Z=Z_f+Z_d$, with

$$Z_f = \exp(-\beta E_0 e^{-t_r} - t_r). \quad (1)$$

representing contribution from the forward-going wave-front, and

$$Z_d = [e^{-Z_0 \exp(-t_r)} - e^{-Z_0}]/Z_0. \quad (2)$$

from all spherical waves diffracted at the edge, respectively. The notation $\beta=1/k_BT$ is used as usual with $k_B$ being the Boltzmann constant and $E_0=h^2/2m\lambda^2$ is the kinetic energy of the particle entering the aperture. The scaled time is defined as $t_r=2ht/ma_0L$. The constant $Z_0$ is defined as the non-zero real solution of the transcend equation $\exp(Z_0)-2Z_0-1=0$. The solution $Z_0$ is irrational and universal. An approximation value $Z_0=1.25643$ is used in numerical calculations here.[13]

Since a pulse of matter wave propagates in space at the speed of group velocity, the time needed for the forward-going wave-front to reach the screen is $D/V_g$, corresponding to a scaled time

$$t_D = 2\lambda D / a_0 L. \quad (3)$$

According to statistical calculations in Ref.[13], if we neglect effects from the aperture on "reflected" waves, the distribution function for probability to find the particle along the $x$-direction can be written as $\rho(x)=\rho_f(x)+\rho_d(x)$ with

$$\rho_f(x) = \frac{\delta(x-D)}{Z}\exp[-2\lambda x/a_0L - \beta E_0 \exp(-2\lambda x/a_0L)] \quad (4.1)$$

$$\rho_d(x) = \frac{\lambda}{Za_0L}\int_0^{(D+x)/2}\frac{\exp[-2\lambda x_1/a_0L - Z_0\exp(-2\lambda x_1/a_0L)]}{D-x_1}dx_1 \quad (4.2)$$

Here $\rho_f(x)$ represents contribution from the forward-going wave-front which is in the form of a $\delta$-function as the wave packet is assumed to be arbitrary sharp. $\rho_d(x)$ is the overall probability projected on $x$-axis, representing contribution from all those waves diffracted at the edge of the forward-going wave-front. To arrive at Eq.(4.2), it is essential that the structure is exactly axial symmetry and the probability of an out-going spherical wave is uniformly distributed on its sphere. Furthermore, it is assumed that diffracted spherical waves at different time are no longer coherent, since the particle can transfer heat with its environment at a finite temperature. The condition of normalization here is

$$\int_{-D}^{D+0^+}\rho(x)dx = 1. \quad (5)$$

Note that $\rho_d(x)$ is nonzero, extending from $x=-D$ to $x=D$, as if a quantum particle would undergo a kind of reflection when it moves freely in between the aperture and the detection screen. In an experiment, particles entering the aperture in an independent way, $\rho_d(x)$ determines thus the residual portion of the stream that may not reach the target within time $t_D$. The ratio of detected particles to incoming particles is equal to the probability for an individual particle to reach the screen. If the number of particles per time entering the aperture is known to be $N_0$, the number of particles per time $N(D)$ detected at the screen is thus given by

$$\frac{N(D)}{N_0} = 1 - \int_{-D}^{D} \rho_d(x) dx$$

$$= \frac{Z_0}{Z_0 + [e^{-Z_0 \exp(-t_D)} - e^{-Z_0}] \exp(\beta E_0 e^{-t_D} + t_D)}. \qquad (6)$$

As an individual particle in the stream is captured by a detector, the thermodynamic process ends and all waves associated with it collapse. As it happened to fail to reach the detector within time $t_D$, whose waves should continue to evolve and effects of the screen on its waves must be taken into account which is beyond the scope of the present work. Such kind of secondary effect can be neglected by assuming that the density of particles in the stream is dilute enough and detection is finished within the time of coherence.

In Fig. 2 we plot the natural logarithm of $N/N_0$ determined by Eq.(6) as a function of $2\lambda D/a_0 L$ for different values of $\beta E_0$. For $\beta E_0 < 4$, the function decreases monotonically with increasing the distance between the aperture and the screen as expected. For larger values of $\beta E_0$, the function shows a different behavior, a local valley and peak structure becomes evident. The result indicates that, in some range away from the aperture, the number of particles detected per time increases if the distance of the screen is increased. This unusual behavior is in some way similar to the oscillatory interference pattern as a function of the diffraction angle in an interferometer. If a particle passed were a classical free particle governed by Newton's laws of motion, it would keep moving with constant speed and we would have the simple result $N(D)/N_0=1$ with no regard to dimension of the hole or the temperature of the surrounding space. In the present work, a quantum particle is described by its matter waves at the de Broglie wavelength. These waves are nonlocal and evolve with time according to Huygens-Fresnel principle. Without the need to ascribe a temperature to the particle, thermodynamic interaction with the surrounding space at a non-zero temperature has been considered within the framework of equilibrium statistical physics. Therefore, Eq.(6) is a manifestation of the wave character of a single particle and is purely quantum mechanical without classical counterpart.

Note also that Eq.(6) relates explicitly the ratio $N(D)/N_0$, which can be measured experimentally, to the factor $\beta E_0$ and the scaled time $2\lambda D/a_0 L$. An experimental verification would measure directly the unknown parameter $L$, which is believed to be temperature dependent for a quantum particle.[12] If $L$ at a temperature is known, thermodynamical quantities such as the time-dependent internal energy and the entropy of the particle in Ref.[13] become to be exact at the temperature. As the result, the motion of a particle at the temperature can be understood for whole process starting from a single quantum state in quantum mechanics to finally in thermal equilibrium with the surrounding space. In such a way, it becomes possible an exact finite-time thermodynamics of an individual quantum particle.

Of special interest is the case when the kinetic energy of an incoming particle is much higher than the thermal energy of the surrounding space. For $\beta E_0 >> 1$, and $t_D << 1$, Eq. (6) is approximately written as

$$\frac{N(D)}{N_0} = \frac{1}{1+\exp(\beta E_0 - Z_0) \times t_D \exp(-\beta E_0 t_D)} . \qquad (7)$$

The local minimum can be easily found to appear at the distance $D_{min}=a_0L/(2\lambda\beta E_0)$, with

$$\frac{N(D_{min})}{N_0} = \frac{\beta E_0}{\beta E_0 + \exp(\beta E_0 - Z_0)} . \qquad (8)$$

Started from the position $D_{min}$, Eq.(6) becomes a increasing function as the distance $D$ is increased. This behavior holds till the position $D_{max}= (a_0L/2\lambda)\ln(\beta E_0)$, at which point the local maximum is reached, with

$$\frac{N(D_{max})}{N_0} = \frac{Z_0}{e\beta E_0(1-e^{-Z_0})} . \qquad (9)$$

What is surprising in the case is that both the minimum and the maximum are simple functions of $\beta E_0$, independent of the unknown parameter $L$.

As the distance between the aperture and the observation screen is large so that $D>>a_0L/2\lambda$, we have

$$\frac{N(D)}{N_0} = \frac{Z_0}{Z_0 + [1-e^{-Z_0}]\exp(2\lambda D/a_0L)}$$

$$\approx \frac{Z_0}{1-e^{-Z_0}}\exp(-2\lambda D/a_0L) . \qquad (10)$$

which decreases with the distance in the simple exponential form. Note that Eq.(10) may only be of theoretical interest, because matter-waves diffracted at the edge of the wave-front expands along the two perpendicular directions at the speed of the group velocity, *i.e.*, it requires physically that the dimension of the screen must be larger than the distance between the aperture and the screen.

Finally, it must be pointed out that an particle here is structureless and the wave packet associated with it is assumed to be arbitrary sharp and effects from its spreading has been neglected. In fact, the wave packet of a particle in experiment is finite and internal structures due to rotational or vibrational freedoms may lead to extra effects such as spreading and emitting of photons.[14] Temperature dependent spreading of the wave packet of a particle and its application to process of decoherence without dissipation have already been discussed before by Ford and O'Connell.[15,16] These effects make things more complicated than believed here, although the decoupling of translational and internal freedoms has been confirmed in quantum interference experiments[6]. Therefore, it is important to use simple atoms and to keep wave packets as sharp as possible in an experiment. For electrons, the length of a wave packet as short as $1\mu m$ can be reached experimentally, while the average distance between two successive particles is as far as 150km.[9]

## 3. Conclusions

In conclusion, we have studied properties of a stream of quantum particles passing through a macroscopic circular aperture of matter-waves within the coherence time of an individual particle. When the density in the stream is dilute enough, the portion of particles detected at the downstream screen shows a valley-peak structure as the distance between the aperture and the screen is increased, depending on both the wavelength of a particle in the stream and the temperature of the environment.

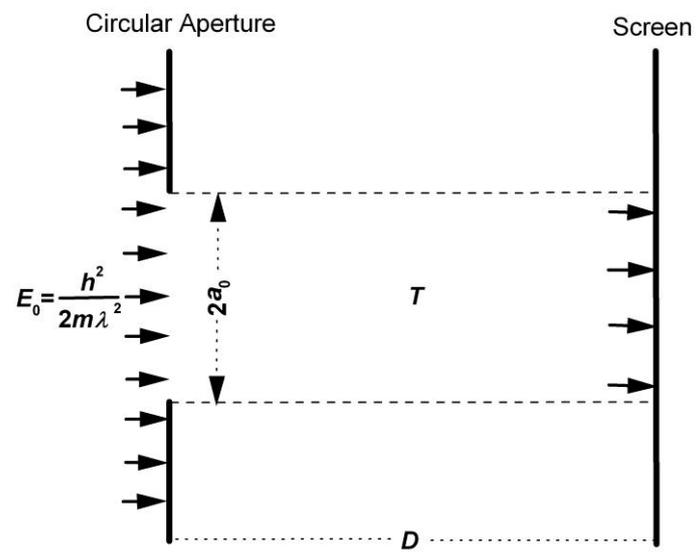

Fig. 1. A stream of free particles with monochromatic wave length pass through a circular matter wave aperture of radius $a_0 \gg \lambda$. Particles reaching the detector screen are counted with single particle resolution. The whole space between the aperture and the screen is at constant temperature $T$.

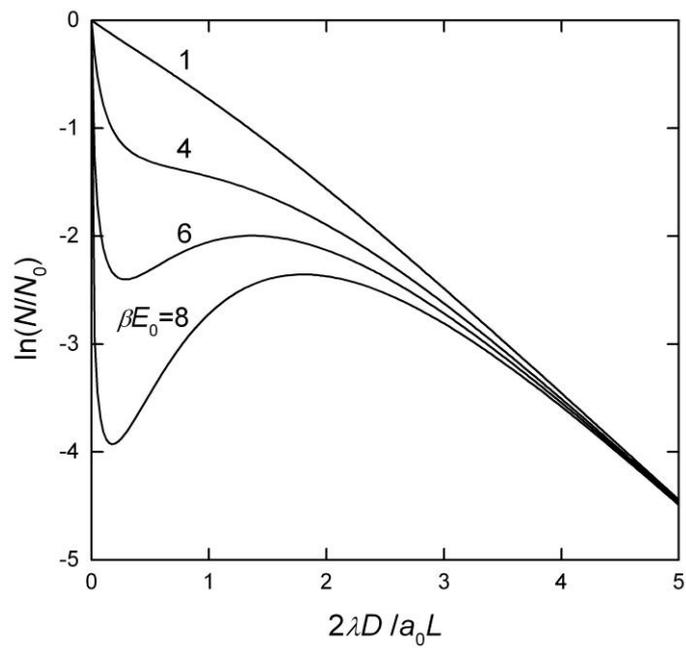

Fig.2. Dependence of the natural logarithm of $N/N_0$ on the distance between the source and the screen, for different values of $\beta E_0$ labeled in the figure.